\documentclass[sigconf]{acmart}

\setcopyright{none}
\renewcommand\footnotetextcopyrightpermission[1]{}

\usepackage{amssymb}
\usepackage{enumitem}

\usepackage{listings}
\newcommand{\pinaff}[1]{%
  \affiliation{%
    \institution{Pinterest, Inc., USA}%
    \country{}}%
  \email{#1}%
}

\begin{document}

\sloppy
\title{Fine-Tuned LLM as a Complementary Predictor Improving Ads System}

\author{Hui Yang}             \pinaff{huiyang@pinterest.com}
\author{Daiwei He}            \pinaff{daiweihe@pinterest.com}
\author{Kevin Jiang}          \pinaff{kevinjiang@pinterest.com}
\author{Taejin Park}          \pinaff{taejinpark@pinterest.com}
\author{Kungang Li}           \pinaff{kungangli@pinterest.com}
\author{Jiajun Luo}           \pinaff{jiajunluo@pinterest.com}
\author{Yuying Chen}          \pinaff{yuyingchen@pinterest.com}
\author{Xinyi Zhang}          \pinaff{xinyizhang@pinterest.com}
\author{Sihan Wang}           \pinaff{sihanwang@pinterest.com}
\author{Haoyu He}             \pinaff{haoyuhe@pinterest.com}
\author{Yu Liu}               \pinaff{lenaliu@pinterest.com}
\author{Lakshmi Manoharan}    \pinaff{lmanoharan@pinterest.com}
\author{David Xue}            \pinaff{davidxue@pinterest.com}
\author{Shubham Barhate}      \pinaff{shubhambarhate@pinterest.com}
\author{Runze Su}             \pinaff{runzesu@pinterest.com}
\author{Duna Zhan}            \pinaff{dzhan@pinterest.com}
\author{Ling Leng}            \pinaff{lleng@pinterest.com}
\author{Siping Ji}            \pinaff{siping@pinterest.com}
\author{Jinfeng Zhuang}       \pinaff{jzhuang@pinterest.com}
\author{Alice Wu}             \pinaff{alicewu@pinterest.com}
\author{Leo Lu}               \pinaff{xlu@pinterest.com}
\author{Han Sun}              \pinaff{hsun@pinterest.com}
\author{Zhifang Liu}          \pinaff{zhifangliu@pinterest.com}
\renewcommand{\shortauthors}{Yang et al.}

\begin{abstract}
Recommendation systems power engagement and monetization across feeds, ads, and short‑video platforms, but translating the latest advances in Large Language Models into Recommendation Systems (RecSys) gains remains rare—particularly in advertising, especially in production‑scale real world industry setups. Prior real‑world LLMs successes typically fall into three buckets: (a) generative retrieval that directly predicts the next items for candidate generation, (b) late‑stage re-ranking that uses LLMs, and (c) auxiliary signal enrichment with LLMs. We introduce a complementary paradigm for ads: a fine‑tuned open‑source LLM used not as a ranker, but as an ads‑specific ancillary predictor — forecasting likely advertisers from user profiles and histories. This LLM‑driven advertiser prediction augments conventional candidate generation and provides informative priors to downstream ranking. Developed in a large-scale production advertising system, our approach produces substantial offline improvements and measurable online business impact, demonstrating that LLM world knowledge and predictive capacity can be efficiently harnessed. Beyond validating LLMs for ads applications, our results show that targeted ancillary predictions can unlock end‑to‑end gains across both retrieval and late‑stage ranking, offering a practical path to LLM‑enhanced recommendation at scale.
\end{abstract}



\maketitle

\section{Introduction}
Recommendation systems are a foundational layer for large-scale consumer platforms, powering feeds, ads, and short‑video experiences. Production systems commonly follow a multi‑stage architecture that balances coverage, latency, and accuracy: an early candidate generation or retrieval stage produces a manageable slate of items, and a downstream ranking stack orders those candidates for user delivery \citep{YouTube2016, DLRM2019}. The retrieval stage emphasizes high recall and efficiency—often via dual‑encoder or two‑tower formulations that map users and items into a shared embedding space for approximate nearest‑neighbor search \citep{TwoTower2020}. The ranking stage is precision‑oriented and typically deploys more expressive but costlier models such as Wide\&Deep, DCN, and deep sparse architectures to capture high‑order cross‑feature interactions and long‑tail effects \citep{WideDeep2016, DCN2017, DLRM2019}. Historically, the dominant improvements in industrial RecSys have stemmed from collaborative‑filtering signals—propagating preferences across similar users and items—augmented by ever‑richer feature crosses and large‑scale representation learning.

Despite the rapid progress in Large Language Models (LLMs), directly transplanting them into production RecSys has proven challenging. First, industrial recommenders rely heavily on sparse, ID‑centric features (user IDs, item/advertiser IDs, campaign/creative IDs), which do not naturally align with token‑based semantic representations \citep{DLRM2019}. Second, modern recommender stacks explicitly model deep cross‑feature interactions and complex calibration objectives; mapping these inductive biases into LLMs remains non‑trivial \citep{DCN2017, WideDeep2016}. Third, LLMs' parameter scales and decoding costs strain real‑time serving budgets at RecSys traffic, where strict tail‑latency SLOs, memory footprints, and cost constraints dominate system design.

Accordingly, the most visible successes of LLMs in recommendation to date have tended to take one of three forms. (a) Generative retrieval: using LLMs to directly produce next‑item or semantic IDs for candidate generation, sometimes via constrained decoding or learned codebooks \citep{GenRet2023, SemanticID2024}. (b) Late‑stage reranking: applying LLMs as powerful, but expensive, rerankers over compact slates \citep{LLMRerank2023}. (c) Auxiliary signal enrichment: prompting or fine‑tuning LLMs to infer side information—item tags, user interests, or hashtags—that can be fed back as features to conventional models \citep{LLMAuxTags2023}. Yet, demonstrable, large‑scale impact in ads recommendation remains relatively rare, where advertiser entities, targeting constraints, and conversion histories play a central role and introduce additional sparsity and mismatch with text‑only semantics.

We propose a complementary paradigm tailored to ads systems: leverage a fine‑tuned open‑source LLM not as a ranker, but as an ads‑specific ancillary predictor that forecasts likely advertisers conditioned on user profiles and conversion histories. Concretely, our method centers on: (1) a structured compilation and prompt template over user profiles and ads‑centric histories (e.g., past conversions, top advertisers/brands/categories/URLs, and demographics); (2) fine‑tuning the LLM to predict next‑advertiser candidates and related auxiliary tags (e.g., user interests) from this structured context; and (3) injecting these predictions into both early retrieval—via advertiser‑targeted candidate filtering—and late‑stage conversion models as additional features. In production, we observe consistent offline gains (e.g., AUC improvements in late‑stage ranking) and measurable online business impact from retrieval‑level targeting guided by the LLM signals.

Our contributions are threefold:
\begin{enumerate}[leftmargin=1.5em]
    \item We identify and address core mismatches between LLM semantics and ads RecSys by formulating an advertiser-prediction task that leverages LLM world knowledge and a conversion layer in between advertiser names and advertiser IDs.
    \item We present a practical recipe---data compilation, prompt design, and fine-tuning---that yields stable signals usable at both retrieval and ranking stages with acceptable latency/cost envelopes.
    \item We demonstrate end-to-end improvements in a large-scale production ads system, showing that compact, LLM-derived ads-oriented predictions complement existing collaborative and feature-engineered signals and unlock gains across the full service pipeline.
\end{enumerate}

Overall, our results suggest a new pragmatic path for LLM‑enhanced recommendation: rather than replacing industrial RecSys stacks or restricting LLMs to reranking, we use fine‑tuned LLMs as complementary predictors that distill heterogeneous histories and world knowledge into actionable advertiser priors—improving both retrieval coverage and ranking precision in ads recommendation at scale.

\section{Related Work}
Recent advances explore how large language models (LLMs) can enhance recommendation quality across retrieval, ranking, and auxiliary prediction. A prominent line of work aligns LLMs with recommendation knowledge through continued pre‑training and task‑specific fine‑tuning. For example, \citet{AlignRec2024} jointly optimize core tasks-retrieval, ranking, and rating prediction together with auxiliary objectives to transfer general language capabilities into recommendation specific reasoning. At the platform scale, \citet{Brew360_2025} report training a foundation LLM predominantly in first‑party behavioral corpora, highlighting the promise of domain‑native pre‑training for utility and safety. Complementarily, instruction‑following formulations cast recommendation as multi‑task alignment over user intents and heterogeneous objectives, demonstrating that a single LLM can adapt to diverse preference elicitation and recommendation formats \citep{RecAsInstr2023}.

Despite these advances, directly deploying a general‑purpose LLM in real‑time industrial stacks remains challenging due to latency, cost, and the prevalence of sparse ID features. Consequently, many successful applications position LLMs at or near the retrieval boundary. Generative retrieval methods produce candidate items via constrained decoding or aligned identifier spaces, showing strong results in large catalogs \citep{EcomGRA_2025, PLUM2025, RecGPT2025}. Related efforts build semantic or structured identifier schemes to bridge token semantics and item IDs, improving retrieval coverage and controllability in production settings \citep{SemanticID2024, GenRet2023}.

A parallel thread borrows the autoregressive sequence‑modeling architecture while eschewing pre‑trained LLMs, focusing purely on ultra‑long user histories and efficient serving. Recent systems scale transducer‑style or decoder‑only models to hundreds of billions (or more) effective parameters for sequential recommendation, targeting extreme‑length contexts and strict tail‑latency budgets \citep{DV3652025, MassiveMem2025}. Others combine sequence models with quantized or learned identifier spaces to reduce vocabulary size and improve generalization \citep{OneRec2025, GenRet2023, SemanticID2024}.

Finally, multi‑objective co‑training that pairs generation with contrastive or retrieval signals has emerged as a practical bridge between LLM reasoning and classical RecSys objectives. NoteLLM \citep{NoteLLM2024} jointly optimizes a note generation task and an item‑to‑item contrastive objective, reporting strong improvements and broad applicability in a large‑scale content platform.

Our work complements these directions by fine‑tuning an open‑source LLM as an ads‑specific ancillary predictor. Rather than replacing retrieval or acting solely as a reranker, we forecast likely advertisers and auxiliary tags from user profiles and conversion histories, and feed these predictions to both early retrieval and late‑stage ranking. This design leverages LLM world knowledge while respecting industrial constraints on latency and sparsity.

\section{Methodology}

\subsection{System Overview}
Our approach uses a fine-tuned LLM as a complementary advertiser predictor rather than a direct end-to-end ads ranker. Given a structured summary of a user's profile and recent behaviors, the model predicts a set of advertisers with high conversion intent together with a short list of user interests. These LLM outputs are then consumed in two ways: (1) as an additional signal for advertiser-targeted candidate generation in ads retrieval, and (2) as featurized inputs to downstream conversion models.

\begin{figure}[t]
    \centering
    \includegraphics[width=0.95\linewidth]{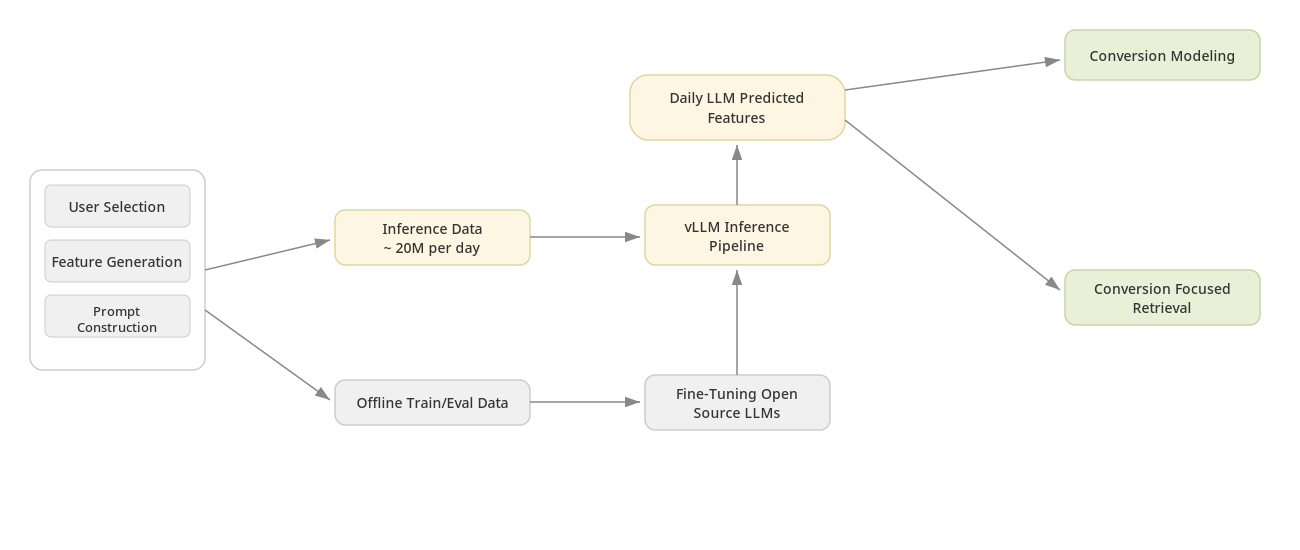}
    \caption{Overall components of the proposed system. The pipeline consists of user selection, feature compilation, prompt-based advertiser prediction, and downstream consumption in retrieval and ranking.}
    \label{fig:overall_components}
\end{figure}

The overall system contains four stages: user selection, feature compilation, prompt-based advertiser prediction, and downstream consumption. User selection constrains daily inference to a high-value segment for which the additional LLM signal is most useful under production latency and cost constraints. Feature compilation converts heterogeneous on-platform and off-platform user activities into a compact textual representation. The LLM is then post-trained to predict advertisers from this structured context. Finally, the predicted advertisers are used as retrieval filters and ranking features in the production ads stack. Figure~\ref{fig:overall_components} illustrates the overall workflow of the proposed system.

\subsection{Data Pipeline}

\subsubsection{User Selection}
For daily inference, we explicitly trade off user coverage against inference cost and speed. The main bottleneck is large-scale daily LLM inference, so we restrict the serving population to active Pinterest users in the U.S. who also have active past conversions within the previous 90 days, including both on-Pinterest and off-Pinterest conversion signals. This design keeps the serving population focused on users with meaningful commercial intent while remaining operationally feasible.

For offline data generation, each example is anchored at date $x$. A user is included if they have valid future conversion labels after $x+1$. Labels are constructed only from advertisers that are active on Pinterest and have active ads spending. In addition, only high-priority conversion types are used as labels.

\subsubsection{Label Construction}
The core prediction target is the first advertiser with which the user converts in the prediction window. In offline evaluation, the label is defined as the first converted advertiser observed in the window from day $x+1$ through day $x+7$. We use the first conversion to align training with a next-advertiser prediction task rather than a multi-label prediction setting. This next-advertiser formulation matches the intended use case of providing advertiser priors for retrieval and ranking.

For training and evaluation, we also maintain distinct user-based train/eval splits to prevent leakage. User IDs are split randomly, and the current split ratio is $0.9/0.1$ for train and evaluation.

\subsubsection{User Feature Compilation}
We compile a structured user representation from both profile features and behavior sequences.

\paragraph{User profile.}
The profile features include age, gender, and user state.

\paragraph{Behavior sequences.}
The behavior summary includes:
\begin{itemize}[leftmargin=1.5em]
    \item onsite searches;
    \item offsite attributed conversions;
    \item offsite matched conversions;
    \item offsite searches related to conversions;
    \item offsite URLs related to conversions;
    \item top categories/interests derived from attributed conversions, matched conversions, onsite searches, and offsite searches;
    \item onsite categories from query data;
    \item offsite categories from URL data;
    \item offsite item brands;
    \item offsite brands extracted from URL data.
\end{itemize}

\paragraph{Advertiser pools.}
We use two advertiser sources in prompt construction:
\begin{itemize}[leftmargin=1.5em]
    \item advertisers with which the user has active past conversions; and
    \item a preset advertiser pool built from daily top-revenue U.S. shopping advertisers (OCPM and ROAS).
\end{itemize}

This representation is designed to preserve recency and commerce relevance while staying within prompt length limits.

\subsection{Prompt Design}
We design stage-specific prompts over the same structured user context. Across all stages, the prompt input contains: (1) active advertisers with past conversions, (2) user profile fields, and (3) behavior summaries including attributed conversions, matched conversions, onsite searches, offsite searches, offsite URLs, top categories, and top brands. The main differences across stages lie in the prediction target, the number of advertisers requested, and the output format. Table~\ref{tab:prompt_stages} summarizes the three stages.

\begin{table}[t]
\centering
\small
\resizebox{\columnwidth}{!}{%
\begin{tabular}{llcl}
\toprule
Stage & Prediction Target & \# Advertisers & Output Format \\
\midrule
SFT & Single next advertiser & 1 & Free-text \\
GRPO & Ranked advertisers + interests & 20 + 5 & Structured XML \\
Inference & Ranked advertisers + interests & 20 + 5 & Structured XML (same as GRPO) \\
\bottomrule
\end{tabular}}
\caption{Summary of stage-specific prompt design. All stages share the same structured user context; they differ in prediction target, advertiser count, and output format.}
\label{tab:prompt_stages}
\end{table}

\paragraph{Post-training prompt for SFT.}
We first perform supervised fine-tuning (SFT) with a simplified prediction target: the model is asked to predict a single next advertiser. The training label is the single next advertiser in the supervision window. Compared with later stages, this formulation focuses the model on the most precise advertiser prediction task while preserving the same structured user context. The SFT prompt is shown in Appendix~\ref{app:sft_prompt}.

\paragraph{Post-training prompt for GRPO.}
After SFT, we further optimize the model with GRPO. In this stage, the model is prompted to generate a ranked list of 20 advertisers together with up to 5 user interests. Although the supervision label remains a single next advertiser, the richer output space provides more informative reward signals during optimization. We choose 20 advertisers instead of 5 because a longer ranked list gives more reward variance per GRPO step. The GRPO prompt is shown in Appendix~\ref{app:grpo_prompt}.

\paragraph{Inference prompt.}
At inference time, we use the same structured user input as in post-training, and the output format is kept the same as in GRPO: the model returns exactly 20 advertisers and up to 5 user interests in XML form. Keeping the inference format aligned with GRPO reduces train--test mismatch and makes the generated outputs directly usable by downstream systems. An example inference output is given in Appendix~\ref{app:inference_output}.

\paragraph{Summary.}
In summary, SFT uses a single-advertiser prompt to learn precise next-advertiser prediction, while GRPO expands the task to ranked advertiser and interest generation under a structured XML output format. Inference follows the same structured output format as GRPO, enabling consistent generation behavior and straightforward downstream consumption.

\subsection{Post-Training}
We study two post-training strategies: supervised fine-tuning (SFT) and GRPO-based reinforcement learning.

\subsubsection{Supervised Fine-Tuning}
For SFT, training focuses on predicting a single next advertiser. The training label is therefore the single next advertiser defined in the label-construction step. This setup directly optimizes the most precise version of the advertiser-prediction task.

\subsubsection{GRPO Training}
For GRPO, the training prompt matches inference: the model predicts 20 advertisers and 5 user interests, while the ground-truth supervision still comes from a single next advertiser. The motivation for predicting 20 advertisers instead of 5 is to produce higher reward variance per GRPO step.

The total reward is defined as
\begin{equation}
R_{\mathrm{total}} = R_{\mathrm{match}} - P_{\mathrm{adv\_len}} - P_{\mathrm{interest\_len}}.
\end{equation}

Here, $R_{\mathrm{match}}$ rewards whether the ground-truth advertiser appears in the predicted advertiser list and where it appears. If the correct advertiser is ranked at position $i$, then
\begin{equation}
R_{\mathrm{match}}(i) = R_{\mathrm{base}}(i) + R_{\mathrm{bonus}}(i),
\end{equation}
where
\begin{equation}
R_{\mathrm{base}}(i) = 0.1 \times (20 - i),
\end{equation}
and
\begin{equation}
R_{\mathrm{bonus}}(i) =
\begin{cases}
2.0, & \text{if } i \leq 4,\\
0, & \text{otherwise.}
\end{cases}
\end{equation}

The length penalty is used for both advertiser count and interest count:
\begin{equation}
P_{\mathrm{len}}(n, n^\ast) =
\begin{cases}
0, & \text{if } n = n^\ast,\\
\min(0.1 \times |n - n^\ast|, 1.0) + 1.0, & \text{otherwise.}
\end{cases}
\end{equation}

This reward encourages both correct advertiser ranking and strict format adherence.

\subsection{Semantic ID (SID) Enhancement}
On top of the existing design, we incorporate Semantic ID (SID) knowledge through continued pre-training (CPT) of the open-source LLM base model, followed by supervised fine-tuning (SFT) for this use case. For offline evaluation, we conducted a side-by-side comparison: the text-only model receives a prompt containing only textual features, whereas the SID-enabled model receives an otherwise identical prompt that additionally includes recent SIDs. The textual content was kept identical across the two settings so that any performance differences can be attributed to the presence or absence of SIDs. For this study, we use a SID with 5 levels and 20248 codes per level constructed from multimodal PinCLIP embeddings \citep{beal2026pinclip} using RQ-VAE \citep{GenRet2023}.

We propose multi-phase LLM post-training approach with SIDs to mitigate catastrophic forgetting issue of general knowledge and instruction following capability:
\begin{itemize}[leftmargin=1.5em]
\item Phase 1: The first phase focuses on a preliminary alignment between the special SID tokens and the normal text tokens space. We curate CPT training data from Pinterest user profile, Pin meta data, user engagement sequences, and SID information (combining Pin title, description, and taxonomy), containing billions of tokens. During this phase, we update only the embeddings of SID tokens while freezing normal text token embeddings and the transformer model parameters.
\item Phase 2: In the second phase, we unfreeze the entire model and run full-parameter pretraining to incorporate recommendation knowledge into the LLM. We curate diverse recommendation datasets with SIDs and mix in a substantial share of general-domain data to further mitigate catastrophic forgetting.
\item Phase 3: After phases 1 and 2, we have obtained an LLM with both SID-based recommendation knowledge and general world knowledge. The third phase is task-specific fine-tuning. For the advertiser-prediction task, we use the same post-training fine-tuning recipe as in the previous section: the model is fine-tuned to predict the next advertiser, with prompts that now include SIDs in addition to text. We can also fine-tune the model to predict next SIDs, which can be used as the candidate retrieval generators and as ranking features.
\end{itemize}

\subsection{Downstream Integration}

\subsubsection{Retrieval: LLM-Based Candidate Generator}
We develop a new candidate generator (CG) that incorporates LLM-predicted advertisers into the ads retrieval stage. Concretely, the predicted advertisers are used as targeting filters, and retrieval under those advertisers is carried out by a two-tower model optimized for user engagement. This CG is designed as a complementary retrieval channel rather than a replacement for the main candidate generators.

The goal of this design is to contribute additional traffic from advertisers with high conversion intent while preserving downstream funnel survival and user experience. In production, this candidate generator is particularly useful for users with weaker existing signal coverage.

\subsubsection{Ranking: LLM-Derived Features}
In addition to retrieval, the LLM-predicted advertisers and user interests are featurized as inputs to downstream conversion models, including ctcvr, vtcvr.

\subsection{Serving and Productionization}
\begin{figure}[t]
    \centering
    \includegraphics[width=0.95\linewidth]{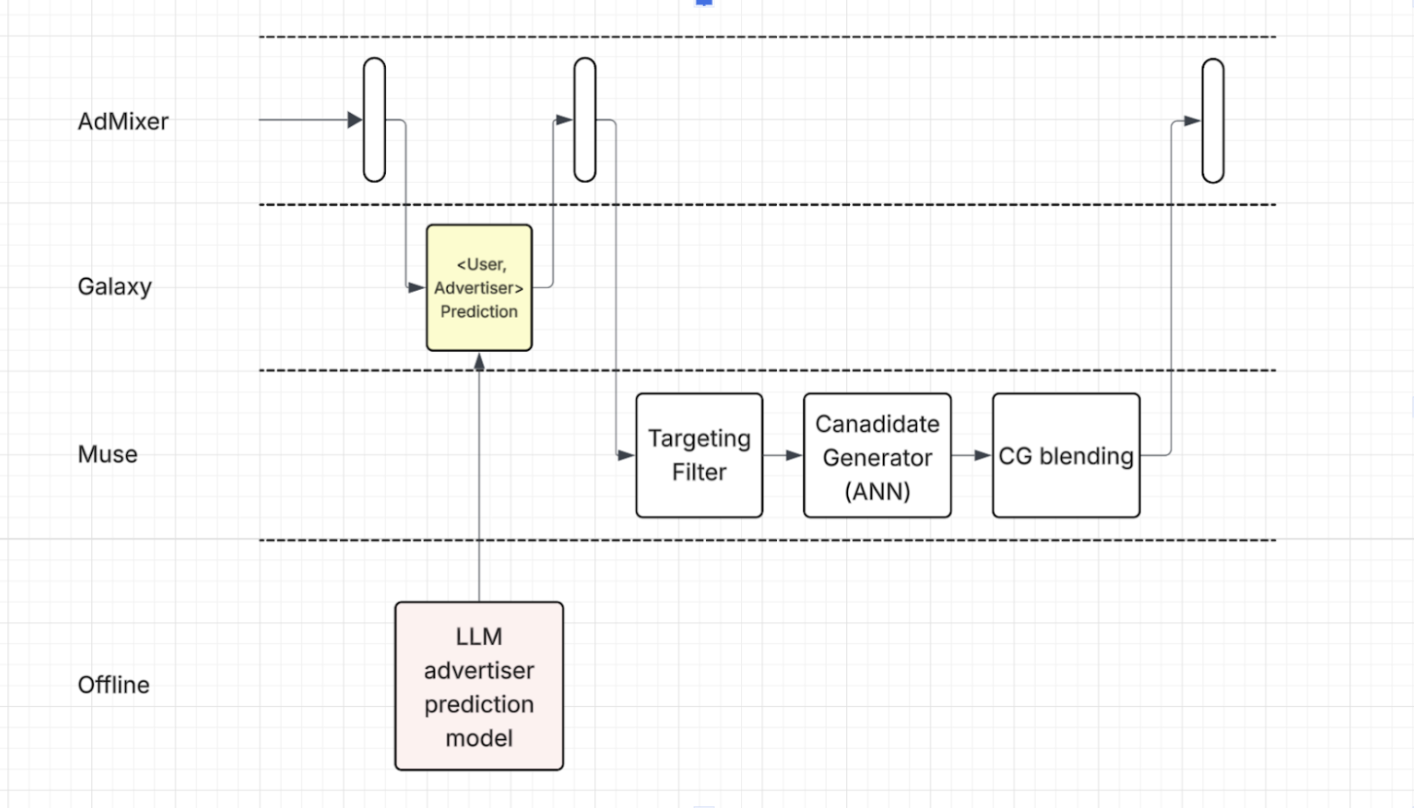}
    \caption{Serving architecture for large-scale inference. The deployment uses a distributed \texttt{vLLM + Ray} stack with batching, prefix caching, fault tolerance, and incremental updates for efficient daily inference.}
    \label{fig:serving_architecture}
\end{figure}

We serve the model with a distributed \texttt{vLLM + Ray} stack for large-scale batch inference. The serving architecture is illustrated in Figure~\ref{fig:serving_architecture}. This setup provides three practical benefits:
\begin{itemize}[leftmargin=1.5em]
    \item prefix caching for repeated prompt templates;
    \item efficient KV-cache management via paged attention;
    \item high GPU utilization through continuous batching.
\end{itemize}

To improve reliability, the inference workflow uses virtual epochs with checkpointing so that failed runs can resume from the latest unfinished epoch rather than restarting. We also perform incremental updates: only users with newly observed activities are re-inferred, which substantially reduces daily inference volume.

\section{Experiments and Results}

\subsection{Experimental Setup}
\label{sec:exp_setup}

The feature snapshot is created at date $x$, and feature ranges are chosen through a trade-off between prompt-length constraints and performance. For example, in our experiments, we use onsite search queries from the last three months and URLs from the last two weeks.

We evaluate on two main data versions:
\begin{itemize}[leftmargin=1.5em]
    \item \textbf{V0:} an experimental dataset containing users with sufficiently long sequence features; specifically, matched-conversion and attributed-conversion sequence lengths are filtered to be at least 10.
    \item \textbf{V1:} a dataset aligned with online serving traffic on date $x$; labels are high-priority conversions from $x+1$ to $x+7$. This setup is closer to production traffic, but has less expressive features.
\end{itemize}

For advertiser prediction, we report recall-based metrics. We also run feature ablations and study the impact of LLM-derived advertisers on downstream conversion models.

We compare zero-shot prompting, SFT, and GRPO-based post-training for advertiser prediction. The main comparisons include: (1) zero-shot prompting on a base open-source model, (2) SFT with different decoding strategies, and (3) GRPO-based training with and without explicit reasoning-oriented prompting.

In addition to standalone advertiser prediction, we evaluate the effect of LLM-predicted advertisers on downstream ads models:
\begin{itemize}[leftmargin=1.5em]
    \item conversion models, including ctcvr and vtcvr;
    \item a production candidate generator in the ads retrieval stage.
\end{itemize}

\subsection{Advertiser Prediction Quality}

\subsubsection{Post-Training Comparisons}

\begin{table}[t]
\centering
\small
\begin{tabular}{lccc}
\toprule
Method & Recall@1 & Recall@5 & Recall@20 \\
\midrule
Zero-shot prompting & 0.346 & 0.567 & 0.655 \\
SFT + beam search & 0.480 & 0.720 & 0.780 \\
SFT + sampling & 0.443 & 0.664 & 0.707 \\
GRPO (reasoning) & 0.496 & 0.684 & 0.766 \\
Zero-shot w/o reasoning & 0.333 & 0.586 & 0.680 \\
GRPO w/o reasoning & 0.490 & 0.688 & 0.778 \\
\bottomrule
\end{tabular}
\caption{Offline advertiser-prediction results on V0.}
\label{tab:v0_results}
\end{table}

Table~\ref{tab:v0_results} shows that all tuned variants outperform zero-shot prompting on V0. In this richer offline setting, SFT with beam search and the no-reasoning variant are the strongest overall among the reported model variants, while SFT with sampling underperforms beam search. More broadly, these results suggest that post-training matters substantially even when the base open-source model already has strong world knowledge.

\begin{table}[t]
\centering
\small
\resizebox{\columnwidth}{!}{%
\begin{tabular}{lccc}
\toprule
Method & Recall@1 & Recall@5 & Recall@20 \\
\midrule
Zero-shot prompting & 0.117 & 0.301 & 0.422 \\
SFT (20-advertiser prompt) & 0.156 & 0.314 & 0.456 \\
SFT (1-advertiser prompt) & 0.214 & 0.413 & 0.501 \\
SFT (1-advertiser prompt) + GRPO & 0.223 & 0.461 & 0.683 \\
SID-enabled SFT (1-advertiser prompt) + GRPO & 0.248 & 0.515  & 0.755 \\
\bottomrule
\end{tabular}}
\caption{Offline advertiser-prediction results on V1.}
\label{tab:v1_results}
\end{table}

V1 is more aligned with online serving traffic and therefore more representative of production conditions. Table~\ref{tab:v1_results} shows a consistent progression from zero-shot prompting to SFT and then to SFT+GRPO. We observe three important learning-stage effects. First, SFT on the 20-advertiser prompt improves performance, but further training on that format can break format-following. Second, SFT on the 1-advertiser prompt mainly improves the most precise recall metric. Third, GRPO on top of the 1-advertiser SFT initialization substantially improves the largest-$K$ recall metric, with a smaller gain on the most precise metric.

The document also notes that explicit reasoning is generally not useful for this task, and that similar overall performance can often be achieved with simpler SFT-style training. This is consistent with the task structure: predicting the next advertiser is closer to preference and intent aggregation than to long-chain reasoning.

\subsubsection{Effect of Semantic ID Enhancement}

The offline evaluation results for the Semantic ID enhancement described in Section~3.5 are presented in Table~\ref{tab:v1_results}. We can see SID knowledge greatly improved the candidate recall rates from the LLM model with more than 10\% improvements across recall@1, 5 and 20. These gains suggest that SIDs provide complementary signal beyond the textual fields already present in the prompt. Intuitively, SIDs encode multimodal content semantics and co‑occurrence structure (via PinCLIP and board‑level curation) that are difficult to reconstruct from advertiser names, query text, and URL strings alone. 

\subsection{Feature Ablation}

\begin{table}[t]
\centering
\small
\begin{tabular}{lc}
\toprule
Ablation & $\Delta$ Recall@5 \\
\midrule
Remove active advertisers with past conversions & -0.1000 \\
Remove offsite URLs & -0.0290 \\
Remove offsite searches & -0.0140 \\
Remove matched conversions & -0.0098 \\
Remove onsite searches & -0.0098 \\
Remove top brands & -0.0079 \\
Remove user profile & -0.0020 \\
Remove attributed conversions & +0.0000 \\
Remove top categories & +0.0039 \\
\midrule
Reorder by performance descending & -0.0049 \\
Reorder + delete negative-performing sequence & -0.0170 \\
Delete negative-performing sequence & -0.0059 \\
\bottomrule
\end{tabular}
\caption{Feature ablation on the zero-shot 4B setup. Negative values indicate performance drops relative to the baseline.}
\label{tab:ablation}
\end{table}

The ablation study in Table~\ref{tab:ablation} shows that \emph{active advertisers with past conversions} are by far the single most important input feature group. Among behavior features, offsite URLs and offsite searches are the next most influential, suggesting that broad off-platform commerce intent carries useful advertiser-level signal. Matched conversions, onsite searches, and top brands also contribute positively. By contrast, user profile features have relatively small marginal impact, and removing top categories slightly improves Recall@5, indicating that this feature can be noisy in the current prompt formulation.

The sequence-ordering experiments further show that naively reordering or deleting behavior sequences does not help, and can substantially hurt performance. This suggests that the original recency-aware compilation is already important to model quality.

\subsection{Downstream Model Impact}

\subsubsection{Conversion Models}

\begin{table}[t]
\centering
\small
\begin{tabular}{lccc}
\toprule
Model & $\Delta$ Loss & $\Delta$ AUC & $\Delta$ PR-AUC \\
\midrule
ctcvr & -0.12\% & +0.06\% & +0.71\% \\
vtcvr & -0.64\% & +0.09\% & +1.64\% \\
\bottomrule
\end{tabular}
\caption{Offline improvement in conversion modeling after adding LLM-derived advertiser features.}
\label{tab:conversion_gains}
\end{table}

Table~\ref{tab:conversion_gains} shows that LLM-predicted advertisers improve both conversion objectives studied in this work. The gains are particularly strong on PR-AUC, which is important in sparse conversion settings.

\subsection{Online Deployment Results}

\subsubsection{Retrieval Experiment Design}
In the online experiment, we deployed the LLM-based candidate generator on all three view types---Home Feed, Related Pins, and Search---for U.S.-market opt-in users only.

We tested multiple training objectives for the retrieval two-tower model:
\begin{itemize}[leftmargin=1.5em]
    \item impressions as positives;
    \item clicks with duration-based weights as positives;
    \item conversions as positives.
\end{itemize}

The experiments show that loss-function choice is critical. Training on impressions improves survival rate but hurts user-engagement metrics. Training directly on conversions is unstable because of label sparsity. Using clicks with duration-based weights provides the best balance between funnel survival and engagement.

We also found that the L0 quota of this candidate generator must be tuned carefully. Because the CG focuses on advertisers with strong conversion intent, its ads tend to receive high CTR/CVR scores and can dominate candidate blending when the quota is too large. This over-concentration later harms advertiser diversity after de-duplication.

\subsubsection{Online Metrics}

\begin{table}[t]
\centering
\small
\begin{tabular}{lc}
\toprule
Metric & Relative Change \\
\midrule
US Shopping Slice - Return on Ad Spend & +4.94\% \\
Opt-in US Shopping Slice - Return on Ad Spend & +6.69\% \\
\bottomrule
\end{tabular}
\caption{Online impact of the LLM-based candidate generator on U.S. Shopping slice}
\label{tab:online_retrieval}
\end{table}

Table~\ref{tab:online_retrieval} shows that within U.S. Shopping Ads, the new candidate generator increased the Return on Ad Spend (RoAS) by 4.94\% (p=0.021) and  by 6.69\% (p=0.029) for the treatment slice (opt-in users).

These online results validate the retrieval use case of the LLM signal. The strongest incremental gains come from the fact that the candidate generator adds traffic from advertisers with high conversion intent that are not sufficiently covered by existing retrieval channels. At the same time, the production findings make clear that the retrieval model objective, CG quota, and target user segment all matter materially for success.

\subsection{Discussion}
The experiments above reveal several cross-cutting observations. First, explicit reasoning is generally not useful for the advertiser-prediction task, and similar overall performance can often be achieved with simpler SFT-style training. This is consistent with the task structure: predicting the next advertiser is closer to preference and intent aggregation than to long-chain reasoning. Second, the LLM-based candidate generator provides incremental value for the opt-in slice, demonstrating that it complements conventional collaborative-filtering signals rather than mimicking them. 

\section{Conclusion}
We presented a practical way to use a fine-tuned LLM as a complementary advertiser predictor in a production ads system. Rather than replacing the existing retrieval and ranking stack, the LLM is used to synthesize heterogeneous user signals into actionable advertiser priors and compact interest summaries. This design is operationally attractive because it provides value at multiple stages of the ads pipeline: it improves advertiser-targeted candidate generation in retrieval and also yields measurable gains when featurized into downstream conversion models.

Our experiments show that post-training is critical for advertiser prediction, that feature design matters substantially, and that the resulting advertiser priors are strong enough to produce both offline and online improvements. In particular, the largest gains come from active past-conversion advertisers and offsite intent signals, while simple profile features contribute relatively less. In production retrieval, the LLM-based candidate generator improves conversion rate and reduces CPA when deployed on an appropriate user segment with carefully tuned blending.

Overall, these results support the paper's central claim: a fine-tuned LLM can serve as a complementary predictor that improves an industrial ads system without requiring the LLM to become the primary ranker. An important direction for future work is the multimodal semantic-ID extension discussed in the project notes; however, that version has not yet been validated online and is therefore left outside the main empirical claims of this paper.

\appendix
\section{Full Prompt Listings}

\subsection{SFT Prompt (Text-based)}
\label{app:sft_prompt}

\begin{lstlisting}
#Role
You are an ad-matching assistant for a commerce/discovery platform. Your task is to predict 1 advertiser most likely to convert and identify user interests based on their profile and behavior.

#Mandatory Requirements
**Advertiser Mining Requirements**
1) Predict 1 advertiser likely to convert using user behavior and profile. Relate to user interests but don't limit to them.
2) Select exactly 1 advertiser from either the Active Advertisers with Past Conversions or the Preset Advertiser List.
3) Ensure diverse interests; cover dominant ones; avoid duplicates within brand groups.
4) Examine URLs for advertiser and item information.
5) Prioritize the most recent data when evaluating signals (weight recency strongly and de-emphasize stale activity).

**Quantity Requirements**
- Provide 1 advertiser.

#Preset Pool
- Preset Advertiser List: {preset_advertiser_pool}

#Input
- Active Advertisers with Past Conversions:
{active_advertisers_with_past_conversions}
- Profile:
Gender: {gender}, Age: {age}, User Type: {userstate}
- Behavior:
Attributed Conversions [{attributed_conversions}]
Matched Conversions [{matched_conversions}]
Onsite Searches [{onsite_searches}]
Offsite Searches [{offsite_searches}]
Offsite URLs [{offsite_urls}]
Top Categories [{top_categories}]
Top Brands [{top_brands}]
\end{lstlisting}

\subsection{SFT Prompt (SID-aware)}
\label{app:sft_sid_prompt}

\begin{lstlisting}
#Role
You are an ad-matching assistant for a commerce/discovery platform. Your task is to predict 1 advertiser most likely to convert and identify user interests based on their profile and behavior.

#Mandatory Requirements
**Advertiser Mining Requirements**
1) Predict 1 advertiser likely to convert using user behavior and profile. Relate to user interests but don't limit to them.
2) Select exactly 1 advertiser from either the Active Advertisers with Past Conversions or the Preset Advertiser List.
3) Ensure diverse interests; cover dominant ones; avoid duplicates within brand groups.
4) Examine URLs for advertiser and item information.
5) Prioritize the most recent data when evaluating signals (weight recency strongly and de-emphasize stale activity).
6) Use the most recent Semantic ID (SID) sequences as the primary intent signal.

**Quantity Requirements**
- Provide 1 advertiser.

#Preset Pool
- Preset Advertiser List: {preset_advertiser_pool}

#Input
- Active Advertisers with Past Conversions:
{active_advertisers_with_past_conversions}
- Profile:
Gender: {gender}, Age: {age}, User Type: {userstate}
- Behavior:
Attributed Conversions [{attributed_conversions}]
Matched Conversions [{matched_conversions}]
Onsite Searches [{onsite_searches}]
Offsite Searches [{offsite_searches}]
Offsite URLs [{offsite_urls}]
Top Categories [{top_categories}]
Top Brands [{top_brands}]
SID Sequences [{sid_sequences}]
\end{lstlisting}

\subsection{GRPO Prompt (Text-based)}
\label{app:grpo_prompt}

\begin{lstlisting}
#Role
You are an ad-matching assistant for a commerce/discovery platform. Your task is to predict top advertisers most likely to convert and identify user interests based on their profile and behavior.

#Mandatory Requirements
**Advertiser Mining Requirements**
1) Predict advertisers likely to convert using user behavior and profile. Relate to user interests but don't limit to them.
2) Select up to 4 advertisers from the Active Advertisers with Past Conversions; choose others from the Preset Advertiser List.
3) Ensure diverse interests; cover dominant ones; avoid duplicates within brand groups.
4) Examine URLs for advertiser and item information.

**Interest Mining Requirements**
1) Group behaviors into user interests; evaluate by recency, frequency, and diversity (R/F/D).
2) Prioritize consistent multi-session patterns; reduce emphasis on short-term spikes.
3) Create a ranked list of user interests from all behavioral signals.
4) Use URLs to find top interests, removing noise from top categories.

**Quantity Requirements**
- Provide up to 5 unique user interests and exactly 20 advertisers.

#Preset Pool
- Preset Advertiser List: {preset_advertiser_pool}

#Output Format
- Return ONLY the following XML.

<answer>
<interests>
[interest 1|interest 2|interest 3|interest 4|interest 5]
</interests>
<advertiser_names>
Advertiser 1|Advertiser 2|Advertiser 3|Advertiser 4|Advertiser 5|Advertiser 6|Advertiser 7|Advertiser 8|Advertiser 9|Advertiser 10|Advertiser 11|Advertiser 12|Advertiser 13|Advertiser 14|Advertiser 15|Advertiser 16|Advertiser 17|Advertiser 18|Advertiser 19|Advertiser 20
</advertiser_names>
</answer>

#Input
- Active Advertisers with Past Conversions:
{active_advertisers_with_past_conversions}
- Profile:
Gender: {gender}, Age: {age}, User Type: {userstate}
- Behavior:
Attributed Conversions [{attributed_conversions}]
Matched Conversions [{matched_conversions}]
Onsite Searches [{onsite_searches}]
Offsite Searches [{offsite_searches}]
Offsite URLs [{offsite_urls}]
Top Categories [{top_categories}]
Top Brands [{top_brands}]
\end{lstlisting}

\subsection{GRPO Prompt (SID-aware)}
\label{app:grpo_sid_prompt}

\begin{lstlisting}
#Role
You are an ad-matching assistant for a commerce/discovery platform. Your task is to predict top advertisers most likely to convert and identify user interests based on their profile and behavior.

#Mandatory Requirements
**Advertiser Mining Requirements**
1) Predict advertisers likely to convert using user behavior and profile. Relate to user interests but don't limit to them.
2) Select up to 4 advertisers from the Active Advertisers with Past Conversions; choose others from the Preset Advertiser List.
3) Ensure diverse interests; cover dominant ones; avoid duplicates within brand groups.
4) Examine URLs for advertiser and item information.
5) Consider the most recent Semantic ID (SID) sequences as a high-priority signal when selecting advertisers.

**Interest Mining Requirements**
1) Group behaviors into user interests; evaluate by recency, frequency, and diversity (R/F/D).
2) Prioritize consistent multi-session patterns; reduce emphasis on short-term spikes.
3) Create a ranked list of user interests from all behavioral signals.
4) Use URLs to find top interests, removing noise from top categories.
5) Use the most recent Semantic ID (SID) sequences as the primary intent signal.

**Quantity Requirements**
- Provide up to 5 unique user interests and exactly 20 advertisers.

#Preset Pool
- Preset Advertiser List: {preset_advertiser_pool}

#Output Format
- Return ONLY the following XML.

<answer>
<interests>
[interest 1|interest 2|interest 3|interest 4|interest 5]
</interests>
<advertiser_names>
Advertiser 1|Advertiser 2|Advertiser 3|Advertiser 4|Advertiser 5|Advertiser 6|Advertiser 7|Advertiser 8|Advertiser 9|Advertiser 10|Advertiser 11|Advertiser 12|Advertiser 13|Advertiser 14|Advertiser 15|Advertiser 16|Advertiser 17|Advertiser 18|Advertiser 19|Advertiser 20
</advertiser_names>
</answer>

#Input
- Active Advertisers with Past Conversions:
{active_advertisers_with_past_conversions}
- Profile:
Gender: {gender}, Age: {age}, User Type: {userstate}
- Behavior:
Attributed Conversions [{attributed_conversions}]
Matched Conversions [{matched_conversions}]
Onsite Searches [{onsite_searches}]
Offsite Searches [{offsite_searches}]
Offsite URLs [{offsite_urls}]
Top Categories [{top_categories}]
Top Brands [{top_brands}]
SID Sequences [{sid_sequences}]
\end{lstlisting}

\subsection{Inference Output Example}
\label{app:inference_output}

\begin{lstlisting}
<answer>
<advertiser_names>
[Advertiser 1|Advertiser 2|Advertiser 3|Advertiser 4|Advertiser 5|Advertiser 6|Advertiser 7|Advertiser 8|Advertiser 9|Advertiser 10|Advertiser 11|Advertiser 12|Advertiser 13|Advertiser 14|Advertiser 15|Advertiser 16|Advertiser 17|Advertiser 18|Advertiser 19|Advertiser 20]
</advertiser_names>
<interests>
[interest 1|interest 2|interest 3|interest 4|interest 5]
</interests>
</answer>
\end{lstlisting}

\bibliographystyle{ACM-Reference-Format}
\bibliography{refs}

\end{document}